\documentclass[twocolumn,pra, aps,superscriptaddress,floatfix]{revtex4-1}

\usepackage{extarrows}
\usepackage{lineno}
\usepackage{mathptmx}
\usepackage{subfigure}
\usepackage{dcolumn}
\usepackage{amsmath,amssymb}
\usepackage{bm}
\usepackage{color}
\usepackage{overpic}
\usepackage{latexsym}
\usepackage{epstopdf}
\usepackage{color}
\usepackage[english]{babel}
\usepackage{latexsym}
\usepackage{stmaryrd}
\usepackage{bigints}
\usepackage{psfrag,graphicx} 
\usepackage{epsf} 
\usepackage{subfigure} 
\usepackage{amsmath} 
\usepackage{amssymb} 
\usepackage{amsfonts}
\usepackage{bm}
\usepackage{natbib}
\usepackage{epstopdf}
\usepackage{appendix}
\usepackage{relsize}

\definecolor{mygrey}{gray}{0.35}
\definecolor{myblue}{rgb}{0.2,0.2,0.8}
\definecolor{myzard}{cmyk}{0,0,0.05,0}
\definecolor{mywhite}{rgb}{1,1,1}
\definecolor{myred}{rgb}{1,0.,0.3}

\usepackage[colorlinks=true,citecolor=myblue,linkcolor=myred]{hyperref}
\def\be{\begin{equation}}
\def\ee{\end{equation}}
\def\ba{\begin{align}}
\def\enda{\end{align}}
\def\bi{\begin{itemize}}
\def\ei{\end{itemize}}

 \def\ee{\mathord{\rm e}}
 
 \def\ii{\mathord{\rm i}}

\def\half{\textstyle\frac{1}{2}}

 \def\ee{\mathord{\rm e}}
 
 \def\ii{\mathord{\rm i}}

\def\half{\textstyle\frac{1}{2}}

\renewcommand{\ii}{{\rm i}}
\renewcommand{\ee}{{\rm e}}

\def\beq{\begin{equation}}
\def\beq{\begin{equation}}
\def\eeq{\end{equation}}

 \newcommand{\ket}[1]{|#1\rangle}
 \newcommand{\bra}[1]{\langle #1|}

\begin{document}


\title{Correlated Chern insulators  in two-dimensional  Raman lattices: \\
a cold-atom  regularization of strongly-coupled four-Fermi field theories }

\author{L. Ziegler}
\affiliation{Departamento de F\'{i}sica Te\'{o}rica, Universidad Complutense, 28040 Madrid, Spain}
\author{E. Tirrito}
\affiliation{International School for Advanced Studies (SISSA), via Bonomea 265, 34136 Trieste, Italy}
\affiliation{ICFO - Institut de Ciencies Fotoniques, The Barcelona Institute of Science and Technology, Av. Carl Friedrich Gauss 3, 08860 Castelldefels (Barcelona), Spain} 
\author{M. Lewenstein}
\affiliation{ICFO - Institut de Ciencies Fotoniques, The Barcelona Institute of Science and Technology, Av. Carl Friedrich Gauss 3, 08860 Castelldefels (Barcelona), Spain} 
\affiliation{ICREA, Lluis Companys 23, 08010 Barcelona, Spain}
\author{S. Hands}
\affiliation{Department of Physics, Faculty of Science and Engineering, Swansea University, Singleton Park, Swansea SA28PP, United Kingdom}
\author{A. Bermudez}
\affiliation{Departamento de F\'{i}sica Te\'{o}rica, Universidad Complutense, 28040 Madrid, Spain}

\begin{abstract}
We show that ultra-cold atoms with synthetic spin-orbit coupling in Raman lattices can be used as  versatile quantum simulators to explore the connections between correlated Chern insulators and strongly-coupled four-Fermi  field theories   related  to the Gross-Neveu model in (2+1) dimensions. Exploiting this multidisciplinary perspective, we  identify a large-$N$ quantum anomalous Hall (QAH) effect  in absence of  any external magnetic field, and use it to delimit regions in parameter space where these correlated topological phases appear, the boundaries of which are controlled by strongly-coupled fixed points of the four-Fermi relativistic field theory. We further show how, for strong interactions, the   QAH effect gives way to magnetic phases described by a two-dimensional quantum compass model in a transverse field.  We present a detailed description of the  phase diagram using the large-$N$ effective potential, and variational techniques such as projected entangled pairs.
\end{abstract}

\maketitle

The   interplay of   quantum mechanics and special relativity leads to the coupling of the intrinsic angular momentum of the electron with its own motion,  underlying the fine structure of atomic spectra~\cite{THOMAS1926}. This so-called {\it spin-orbit coupling} (SOC) can be  accounted for by  the Dirac equation~\cite{10.2307/94981,10.2307/95359} and, ultimately, by quantum electrodynamics~\cite{10.2307/j.ctv10crg18}. Indeed,  it is the comparison of perturbative calculations of this weakly-coupled   quantum field theory (QFT)~\cite{PhysRev.73.416,PhysRevLett.109.111807}  with high-precision measurements of the intrinsic magnetic moment of the electron~\cite{PhysRevLett.100.120801} and  the fine structure constant~\cite{Parker191}, which yields  the most accurate tests of this QFT.  
              
              SOC has also become a  cornerstone of modern condensed matter~\cite{Bychkov_1984,PhysRev.100.580}, as it underlies the experimental discovery~\cite{Konig766,Hsieh2008} of  a   mechanism for the ordering of matter~\cite{PhysRevLett.95.146802,PhysRevLett.95.226801,Bernevig1757} characterised by the unique role of topology~\cite{RevModPhys.82.3045,RevModPhys.83.1057,classification_spt}. This is    epitomised by  Chern insulators~\cite{PhysRevLett.61.2015,PhysRevB.74.085308,PhysRevB.78.195424}, which have  an insulating bulk with  a non-zero topological invariant    and, simultaneously, host current-carrying states at the boundaries. It is remarkable that  elaborate concepts of algebraic topology~\cite{nakahara_2017}, such as the first Chern number of a fibre bundle~\cite{10.2307/1969037}, emerge naturally  from the band structure of these materials and, more importantly, that they become responsible for the robust quantisation of the transverse conductivity even  without any external magnetic field, i.e. a quantum anomalous Hall  effect~\cite{RevModPhys.89.040502,doi:10.1146/annurev-conmatphys-031115-011417}. Besides their fundamental interest,  topological materials can  be used to build  devices with novel functionalities and promising technological  applications~\cite{Soumyanarayanan2016}.
              
              These SOC phenomena might seem rather simple at first sight,  as topological insulators  employ   band-structure methods based on free electrons, whereas calculations for this QFT  rely on perturbation theory about a weakly-coupled fixed point. Nonetheless, this simplicity is deceiving, as numerous  open questions  arise as one departs from these limits,  exploring  {\it (i)} topological phases in the presence of strong SOC and inter-particle interactions that can lead, for instance, to {\it correlated Chern insulators} and phase transitions thereof, or {\it (ii)}  relativistic  QFTs with SOC that are controlled by  strongly-coupled fixed points, the properties of which can only be accessed non-perturbatively. As argued here, these seemingly unrelated topics can be    connected by considering SOC in  light of specific  discretizations~\cite{Wilson1977} of {\it strongly-coupled four-Fermi  QFTs}~\cite{HANDS199329,hep-lat/9706018,Braun_2012}. Originally introduced by Enrico Fermi in the context of the weak nuclear interactions~\cite{Fermi1934},  four-Fermi QFTs such as the Nambu-Jona-Lasinio~\cite{PhysRev.122.345} or  Gross-Neveu~\cite{PhysRevD.10.3235} models are nowadays considered as effective QFTs that capture the essence   of dynamical chiral symmetry breaking, a phenomenon relevant to the  strong nuclear interactions. We show that,   dispensing with the notion of chirality in  the context of   a  2D Hubbard model~\cite{PRSLSA_276_238} with strong SOC, analogous strongly-coupled four-Fermi QFTs emerge in the long-wavelength limit of correlated Chern insulators. 
              
              This connection can be leveraged to develop  non-perturbative studies of  Chern insulators with interacting electrons following two strategies. On the one hand, one can interpret the lattice as a mere regularisation~\cite{PhysRevD.10.2445}, and exploit the  machinery of  lattice field theories~\cite{gattringer_lang_2010} to explore correlation effects, understanding the nature of the interaction-driven quantum phase transitions  by connecting them to strongly-coupled fixed points. On the other hand, the connection to SOC suggests a different type of regularization. Rather than using the lattice as an artificial  scaffolding for numerical computations, one may consider experimental setups, where this lattice is indeed physical, and both the SOC and four-Fermi interactions appear naturally. As discussed below, Fermi gases with synthetic SOC in optical lattices are a particularly-promising platform~\cite{Galitski2013,book_soc}, as they can provide a cold-atom regularisation of strongly-coupled  QFTs and serve as quantum simulators~\cite{Feynman_1982,doi:10.1080/00018730701223200,Bloch2012}  of  correlated Chern insulators. Considering the shortage of topological materials where correlations have been shown to play a  role, we believe that our results are encouraging,   identifying cold atoms as a viable platform to explore the intricate connections of interacting topological matter and strongly-coupled relativistic QFTs.

	      \textit{The model.--} We consider a Wilsonian
discretization~\cite{Wilson1977} of a Hamiltonian QFT for $N$
flavours of self-interacting Dirac fields $\{\psi_c(\boldsymbol{x})\}_{c=1}^N$ which, in natural units, 
reads
\begin{widetext} 
\beq \label{eq:H_lattice}
H=a_1a_2\!\!\sum_{\boldsymbol{x}\in\Lambda_{\rm
s}}\left[\sum_{j=1,2}\!\!\left(\!-\overline{\Psi}(\boldsymbol{x})\left(\frac{\ii\gamma^j}{2a_j}+\frac{r_j}{2a_j}\right)\Psi(\boldsymbol{x}+a_j\textbf{e}_j)+\overline{\Psi}(\boldsymbol{x})\left(\frac{m}{4}+\frac{r_j}{2a_j}\right)\Psi(\boldsymbol{x})+{\rm
H.c.}\right)-\frac{g^2}{2N}\bigg(\!\overline{\Psi}(\boldsymbol{x}){\Psi}(\boldsymbol{x})\!\!\bigg)^{\!\!2}\right],
\eeq 
\end{widetext} 
where we introduced
$\Psi(\boldsymbol{x})=(\psi_1(\boldsymbol{x}),\cdots,\psi_N(\boldsymbol{x}))^{\rm
t}$,
$\overline{\Psi}(\boldsymbol{x})=({\psi}^\dagger_1(\boldsymbol{x})\gamma^0,\cdots,{\psi}^\dagger_N(\boldsymbol{x})\gamma^0)$
is the adjoint,
and $\gamma^0,\gamma^1,\gamma^2$ are the so-called gamma matrices for a
(2+1)-dimensional  spacetime
$\{\gamma^\mu,\gamma^\nu\}=2g^{\mu\nu}\mathbb{I}$, where
$g^{\mu\nu}={\rm diag}(1,-1,-1)$ is   Minkowski's metric with spacetime indexes
$\mu,\nu\in\{0,1,2\}$.
  
   Inspired by the approach of  Hamiltonian lattice field
theories~\cite{PhysRevD.11.395}, in which only the spatial coordinates are
discretised,  we  consider  an anisotropic rectangular lattice $\Lambda_{\rm
s}=\{\boldsymbol{x}=n_1a_1\textbf{e}_1+n_2a_2\textbf{e}_2, \hspace{0.5ex} \forall \hspace{0.5ex}  (n_1,n_2)\in\mathbb{Z}_{N_1}\times\mathbb{Z}_{N_2}\}$ 
with   lattice spacings $a_1,a_2$ along the Cartesian unit vectors $\textbf{e}_1,\textbf{e}_2$. 
In Eq.~\eqref{eq:H_lattice}, we have also introduced the Wilson parameters $\{r_j\}$, a bare mass  $m$, and a
coupling  $g^2$  that controls the strength of the simplest  
fermion-fermion interaction with Lorentz invariance.
 In the long-wavelength limit, one obtains   $4$
fermion doublers~\cite{NIELSEN198120,NIELSEN1981173}, each    described
by a continuum QFT for an $N$-flavour Dirac spinor $\{\Psi_{\boldsymbol{n}_{
d}}(\boldsymbol{x})\}_{\boldsymbol{n}_{ d}}$,  $\boldsymbol{n}_{
d}=(n_{d,1},n_{d,2})\in\{0,1\}\times\{0,1\}$ with a different
bare mass \beq \label{eq:masses} m_{\boldsymbol{n}_{
d}}=m+2n_{{d},1}r_1/a_1+2n_{{d},2}r_2/a_2.  \eeq We note that the fermion doublers
will interact with each other as soon as  the four-Fermi coupling is switched
on,  playing a key role in the properties of the correlated Chern insulators.
  
  Let us discuss the connection of Eq.~\eqref{eq:H_lattice} to SOC. In  (2+1) dimensions, one can choose $2\times2$  gamma matrices, such that 
the  spinor representation of rotations around the normal vector of the  plane leads to field components
$\psi^{\phantom{\dagger}}_{1}\!(\boldsymbol{x}), \psi^{\phantom{\dagger}}_{2}\!(\boldsymbol{x})$  that can be identified with    spin up/down fermions. If one chooses  
$\gamma^0=\sigma^z$,$\gamma^1=\ii\sigma^x$, $\gamma^2=-\ii\sigma^y$, the  spin-flip  tunnelings in Eq.~\eqref{eq:H_lattice}, ${\Psi}^\dagger(\boldsymbol{x})\frac{\ii\sigma^x}{2a_2}\Psi(\boldsymbol{x}+a_2\textbf{e}_2)-{\Psi}^\dagger(\boldsymbol{x})\frac{\ii\sigma^y}{2a_1}\Psi(\boldsymbol{x}+a_1\textbf{e}_1)$, can be understood as the  second-quantised lattice version  of  Rashba SOC ${\bf e}_3\cdot(\boldsymbol{p}\wedge\boldsymbol{\sigma})=\ii\sigma^x\partial_y-\ii\sigma^y\partial_x$~\cite{Bychkov_1984}. From this perspective, the  combination of fermion bilinears~\eqref{eq:H_lattice} would correspond to a Dirac-type SOC~\cite{book_soc} although, in view of the underlying discretisation~\cite{Wilson1977}, it might be more appropriate to refer to it as a {\it Dirac-Wilson SOC}. In contrast to  (2+1)  Gross-Neveu models~\cite{HANDS199329}, in which a discrete chiral
symmetry   $\Psi(\boldsymbol{x})\to\gamma^5\Psi(\boldsymbol{x})$  can be enforced using 4-component 
spinors,   
$\gamma^5=\ii\gamma^0\gamma^1\gamma^2$  is simply the identity in our case, such that we  can  dispense with the notion of chiral symmetry.  As discussed below, this  has important
consequences for the nature of the phases of Eq.~\eqref{eq:H_lattice}.

		 \textit{Cold-atom quantum simulator.--} We
now describe a mapping of the couplings of the  single-flavour  Hamiltonian~\eqref{eq:H_lattice}, where we choose the standard Dirac basis $\gamma^0=\sigma^z$,$\gamma^1=\ii\sigma^y$, $\gamma^2=-\ii\sigma^x$, to the  experimental parameters of ultra-cold  atoms  in {\it Raman optical lattices}~\cite{Wu83,PhysRevLett.121.150401,Songeaao4748}. To   get a fully-tunable quantum simulator of Eq.~\eqref{eq:H_lattice} in this basis, we shall generalise the synthetic  SOC scheme of~\cite{PhysRevLett.112.086401,PhysRevLett.113.059901}, such that  atoms in two  states $\{\ket{{\uparrow}},\ket{{\downarrow}}\}$ are subjected to the  periodic  potential 
\beq
\label{eq:raman_potential}
\begin{split}
V=&\frac{V_{0,1}}{2}\cos^2\!(kx_1)\mathbb{I}_2+\frac{\tilde{V}_{0,1}}{2}\cos kx_1\ee^{\ii(kx_2-\Delta\omega t-\phi_1)}\sigma^+\\
+&\frac{V_{0,2}}{2}\cos^2\!(kx_2)\mathbb{I}_2+\frac{\tilde{V}_{0,2}}{2}\cos kx_2\ee^{\ii(kx_1-\Delta\omega t-\phi_2)}\sigma^++{\rm H.c.}.
\end{split}
\eeq
Here, $V_{0,j}$ ($\tilde{V}_{0,j}$) stem from  ac-Stark shifts (Rabi frequencies) of  pairs of counter-propagating (orthogonal) laser beams with wavelength $\lambda=2\pi/k$. The Raman term $\tilde{V}_{0,1} (\tilde{V}_{0,2})$ with    relative phase $\phi_1$ ($\phi_2$) induces a two-photon transition between the internal states $\sigma^+=\ket{{\uparrow}}\!\!\bra{{\downarrow}}$  when the beatnote of the lasers is 
$\Delta\omega=\omega_0-\delta$, where $\delta\ll\omega_0$ is the detuning. Accordingly,  the atom absorbs a photon from the standing wave along  $x_1$ ($x_2$), and emits it in the orthogonal travelling wave  along $x_2$ ($x_1$).  To minimise  residual photon scattering and heating, one may consider lanthanide~\cite{soc_lanthanides,soc_lanthanides_bis,soc_clock} or alkali-earth~\cite{soc_clock_bis,soc_clock_dynamics} atoms.

As customary for ultra-cold atoms in optical lattices~\cite{PhysRevLett.81.3108,PhysRevLett.89.220407,RevModPhys.80.885}, in the regime of deep potentials $V_{0,j},\tilde{V}_{0,j}\gg E_{\rm R}=k^2/2m$, where $m$ is the mass of the atoms, the dynamics of the  gas can be expressed in terms of a lattice model where the atoms tunnel between neighbouring minima of the potential, and collide in pairs with an $s$-wave scattering length $a_s$. The  specific interference pattern in Eq.~\eqref{eq:raman_potential} is crucial, as it ensures that the Raman terms do not contribute with on-site spin flips, but  drive instead spin-flip tunnelings along the two spatial directions with a tunable relative phase.  Using a basis of Wannier functions in the single-band approximation, and working in a rotating frame with respect to the Raman terms, one finds
\beq
\label{eq:cold_atom_h}
\begin{split}
H=&-\sum_{\boldsymbol{n},j}\left(t_{\! j}^{\phantom{\dagger}}\!\!\left({f}^\dagger_{\boldsymbol{n},\uparrow}{f}^{\phantom{\dagger}}_{\boldsymbol{n}+{\bf e}_j,\uparrow}+{f}^\dagger_{\boldsymbol{n},\downarrow}{f}^{\phantom{\dagger}}_{\boldsymbol{n}+{\bf e}_j,\downarrow}\right)+{\rm H.c.}\right)\\
&-\sum_{\boldsymbol{n},j}\left(\ii \tilde{t}_{\! j}^{\phantom{\dagger}}\ee^{-\ii\phi_{j,\boldsymbol{n}}}\!\!\left({f}^\dagger_{\boldsymbol{n},\uparrow}{f}^{\phantom{\dagger}}_{\boldsymbol{n}+{\bf e}_j,\downarrow}-{f}^\dagger_{\boldsymbol{n},\uparrow}{f}^{\phantom{\dagger}}_{\boldsymbol{n}-{\bf e}_j,\downarrow}\right)+{\rm H.c.}\right)\\
&+\sum_{\boldsymbol{n}}U_{\uparrow\downarrow}^{\phantom{\dagger}}{f}^\dagger_{\boldsymbol{n},\uparrow}{f}^\dagger_{\boldsymbol{n},\downarrow}{f}^{\phantom{\dagger}}_{\boldsymbol{n},\downarrow}{f}^{\phantom{\dagger}}_{\boldsymbol{n},\uparrow}+\frac{\delta}{2}\!\!\left({f}^\dagger_{\boldsymbol{n},\uparrow}{f}^{\phantom{\dagger}}_{\boldsymbol{n},\uparrow}-{f}^\dagger_{\boldsymbol{n},\downarrow}{f}^{\phantom{\dagger}}_{\boldsymbol{n},\downarrow}\right)\!,
\end{split}
\eeq
where the fermionic  operators ${f}^{{\dagger}}_{\boldsymbol{n},s}({f}^{\phantom{\dagger}}_{\boldsymbol{n},s})$  create (annihilate) an atom  in state $s\in\{\uparrow,\downarrow\}$ at a minimum of the  ac-Stark shifts of Eq.~\eqref{eq:raman_potential}, namely $\boldsymbol{x}_{\boldsymbol{n}}^0=\sum_j(\frac{\lambda}{4}+\frac{\lambda}{2} n_j){\bf e}_j$. Therefore, the terms above correspond to  spin-independent tunnelings of strength $t_j=4(E_{\rm R}/\sqrt{\pi})(V_{0,j}/E_{\rm R})^{3/4}\ee^{-2\sqrt{V_{0,j}/E_{\rm R}}}$, and  contact Hubbard interactions of strength $U_{\uparrow\downarrow}=\sqrt{8/\pi}k_0a_s E_{\rm R}(V_{0,1}V_{0,2}/E_{\rm R}^2)^{1/4}$. Using a  Gaussian approximation for the Wannier functions, we obtain a spin-flip tunneling along the $x_j$-axis with  strength $\tilde{t}_j=\tilde{V}_{0,j}\ee^{-(\pi^2/4)\sqrt{V_{0,j}/E_{\rm R}}}$
             and phase $\phi_{j,\boldsymbol{n}}=\phi_{j}-\pi(n_1+n_2)$. 
             
             In order to find the mapping between the cold-atom model~\eqref{eq:cold_atom_h} and the four-Fermi lattice field theory~\eqref{eq:H_lattice} for $N=1$, we must rescale the atomic operators such that they have the correct field units, and perform a gauge transformation $\psi^{\phantom{\dagger}}_{1,\uparrow}\!(\boldsymbol{x}_{\boldsymbol{n}}^0)={f}^{\phantom{\dagger}}_{\boldsymbol{n},\uparrow}/\sqrt{a_1 a_2}, \psi^{\phantom{\dagger}}_{1,\downarrow}\!(\boldsymbol{x}_{\boldsymbol{n}}^0)=\ee^{\ii\pi(n_1+n_2)}{f}^{\phantom{\dagger}}_{\boldsymbol{n},\uparrow}/\sqrt{a_1 a_2}$. Setting the Raman-beam  phases  to $\phi_1=0,\phi_2=\pi/2$, we find that  model~\eqref{eq:cold_atom_h} maps directly to the  lattice field theory~\eqref{eq:H_lattice} with
             \beq
            \label{eq:parameters} 
             a_j=\frac{1}{2\tilde{t}_j},\hspace{2ex} r_j=\frac{t_j}{\tilde{t}_j},\hspace{2ex} m=\frac{\delta}{2}-2(t_1+t_2),\hspace{2ex}g^2 =\frac{U_{\uparrow\downarrow}}{4\tilde{t}_1\tilde{t}_2}.
             \eeq
             Let us note the following important point. The expression above relates the anisotropic lattice spacings of the four-Fermi  field theory~\eqref{eq:H_lattice} with the atom tunneling strengths, not with the distance $\lambda/2$ between neighbouring minima of the optical lattice. 
             Therefore, taking the continuum limit in the lattice field theory~\eqref{eq:H_lattice} does not require sending the laser wavelength $\lambda\to 0$, but instead setting the experimental parameters $(t_j,\tilde{t}_{j},\delta,U_{\uparrow\downarrow})$ to certain values, such that the bare couplings $(m, a_j,r_j, g^2)$ lie in the vicinity of a critical point. Here,   the energy gap is much smaller than the tunnelings $\Delta\epsilon\ll \tilde{t}_j$, and the relevant length scale   $\xi\gg a_1,a_2$ leads to a continuum QFT.

                \textit{Large-$N$ quantum anomalous Hall effect.--} In the non-interacting  $g^2=0$ and isotropic  $a_1=a_2=:a, r_1=r_2=1$ limits, the  single-flavour Hamiltonian~\eqref{eq:H_lattice}  corresponds to  the square-lattice version~\cite{PhysRevB.74.085308,PhysRevB.78.195424} of Haldane's model~\cite{PhysRevLett.61.2015,haldane_esslinger} of the quantum anomalous Hall (QAH) effect~\cite{doi:10.1146/annurev-conmatphys-031115-011417}. This QAH effect has been realised in  semiconducting ferromagnetic  films
                ~\cite{Chang167,Chang2015}, observing a quantised  conductance transverse to a bias potential  for vanishing magnetic fields. Regarding Eq.~\eqref{eq:H_lattice}, this quantisation depends on the  Chern number via
                \beq
                \label{eq:Chern_free}
                \sigma_{xy}=\frac{e^2}{h}N_{\rm Ch}, \hspace{2ex} N_{\rm Ch}=\frac{N}{2}\sum_{\boldsymbol{n}_d}(-1)^{(n_{d,1}+n_{d,2})}{\rm sign}(m_{\boldsymbol{n}_d}).
                \eeq
                According to the doubler masses~\eqref{eq:masses}, and using Eq.~\eqref{eq:Chern_free}
  in the single-flavour limit $N=1$,  one finds that the Chern number is
quantised to a non-zero integer $N_{\rm Ch}=-1$  when $ma\in(-2,0)$,  whereas
$N_{\rm Ch}=+1$ when  $ma\in(-4,-2)$, both of which lead to a QAH phase, i.e. a Chern insulator, and $N_{\rm Ch}=0$ otherwise.   
Let us note, however, that  the  spinor components  correspond to bonding/anti-bonding orbitals of the upper/lower surface
states of these semiconducting thin films~\cite{Chang167,Chang2015}.  Correlation effects, which do not seem to play any role in these thin films~\cite{Chang167,Chang2015} would not be described, in any case,     by
the Lorentz-invariant four-Fermi term~\eqref{eq:H_lattice}. Instead, for 
cold atoms with the Dirac Wilson SOC~\eqref{eq:cold_atom_h}, the spinor is formed by     two
 internal states that  interact naturally via such a four-Fermi term. The key advantage of the cold-atom
realisation hereby proposed is that  the coupling
strength $g^2$, as well as all other microscopic parameters~\eqref{eq:parameters}, can be experimentally
tuned. This brings a unique opportunity to realise correlated Chern insulators
with a neat connection to strongly-coupled  QFTs originally considered
in high-energy physics. In this 
context~\cite{PhysRevLett.105.190404,PhysRevLett.108.181807,PhysRevX.7.031057,
BERMUDEZ2018149,kuno,PhysRevB.99.125106,verstraete_GN}, topological insulators in different
symmetry classes and dimensions~\cite{Ryu_2010} correspond to lower-dimensional versions of the so-called domain-wall
fermions~\cite{KAPLAN1992342}, where the topological invariants control  a
Chern-Simons-type response to  external gauge
fields~\cite{PhysRevB.78.195424,10.1143/PTP.73.528,GOLTERMAN1993219,2008.01743}.
                
		Let us now describe the fate of these Chern insulators and the QAH effect as we switch on interactions  $g^2>0$. From the analogy with the Gross-Neveu (GN)
model~\cite{PhysRevD.10.3235}, one expects that the phenomenon of dynamical mass
generation by the formation of a homogeneous scalar condensate $\Sigma_0\propto\langle
\overline{\Psi}(\boldsymbol{x})\Psi(\boldsymbol{x})\rangle$ will play an
important role. We note, however, that due to the particular representation of
the gamma matrices, this condensate will not be associated to chiral symmetry
breaking as would occur for the  (1+1) version~\cite{PhysRevD.10.3235,PhysRevX.7.031057,
BERMUDEZ2018149,kuno,PhysRevB.99.125106,verstraete_GN}, or for (2+1) models with a non-trivial $\gamma^5$
matrix~\cite{HANDS199329,hep-lat/9706018}. If such models are discretised
following the Wilson prescription, the fermions can also form a pseudo-scalar
condensate $\Pi_0\propto\langle
\overline{\Psi}(\boldsymbol{x})\ii\gamma^5\Psi(\boldsymbol{x})\rangle$ through
the spontaneous breakdown of parity
$\Psi(\boldsymbol{x})\to\gamma^0\Psi(-\boldsymbol{x})$~\cite{PhysRevD.30.2653,PhysRevD.58.074501}.
In contrast, we identify two different $\pi$-condensates $\Pi_1\propto\langle
\overline{\Psi}(\boldsymbol{x})\gamma^1\Psi(\boldsymbol{x})\rangle$ and
$\Pi_2\propto\langle
\overline{\Psi}(\boldsymbol{x})\gamma^2\Psi(\boldsymbol{x})\rangle$, the
non-zero values of which also lead to the spontaneous breakdown of the discrete
parity symmetry and, additionally, to the breakdown of Lorentz-invariance  even in the
long-wavelength limit. In the language of the underlying 
Hubbard model with SOC~\eqref{eq:cold_atom_h}, these $\pi$-condensates correspond to two orthogonal  ferromagnetic orders, namely  $\Pi_1\propto\langle{f}^{{\dagger}}_{\boldsymbol{n},\uparrow}{f}^{\phantom{\dagger}}_{\boldsymbol{n},\downarrow} \rangle+\langle{f}^{{\dagger}}_{\boldsymbol{n},\downarrow}{f}^{\phantom{\dagger}}_{\boldsymbol{n},\uparrow} \rangle$, and
$\Pi_2\propto\ii\langle{f}^{{\dagger}}_{\boldsymbol{n},\uparrow}{f}^{\phantom{\dagger}}_{\boldsymbol{n},\downarrow} \rangle-\ii\langle{f}^{{\dagger}}_{\boldsymbol{n},\downarrow}{f}^{\phantom{\dagger}}_{\boldsymbol{n},\uparrow} \rangle$, 
which break the combined $\mathbb{Z}_2$ spin and lattice inversions.
                
	       In order to understand how  these condensates affect the Chern
insulator and QAH phases discussed above, we make use of non-perturbative {\it 
large-$N$ techniques} developed in the context of  QFTs in Euclidean spacetime $x=(\ii t,\boldsymbol{x})$~\cite{coleman_1985}. Let us outline our calculation. First of all, we
introduce three auxiliary scalar fields $\sigma(x),\pi_1(x),\pi_2(x)$, which do
not propagate, but act instead as mediators of the contact four-Fermi term~\eqref{eq:H_lattice}. This 
allows us to
organise the Feynman diagrams of the QFT according to their order in $N$, and identify the
leading contribution when $N\to\infty$, which contains  diagrams with a
single fermion loop and an even number of external  lines  for  the auxiliary fields. In this way,
we can obtain analytically the radiative corrections to the classical
potential, which contain the leading-order
contribution~\cite{PhysRevD.10.2491,PhysRevD.10.3235} to the 
 {\it effective
potential}~\cite{PhysRevD.7.1888}. 
In our case,
we are interested in the effective potential  for 
$\boldsymbol{\pi}({x})=({\pi}_1({x}),{\pi}_2({x}))$,  the
minimum of which
 can be used to locate the spontaneous  breakdown of parity $\boldsymbol{\Pi}:=(\Pi_1,\Pi_2)=\langle\boldsymbol{\pi}({x})\rangle \neq\boldsymbol{0}$, $\forall
x$.  
By resumming the  diagrams
to all orders of the coupling strength, we obtain

\beq \begin{split}
\label{eq:v_eff} {V_{\rm eff}(\boldsymbol{\tilde{\Pi}})\over N}&=\frac{\boldsymbol{\tilde{\Pi}}^2}{2\tilde{g}^2}
-\sum_{{k}}\!\log\!\!\left(\!1+\frac{\sum_\mu\tilde{\Pi}_\mu^2}{M_{{k}}^2+\sum_\mu
p_{\mu,{k}}^2}\!\right)\phantom{,} \\ &
-\frac{1}{2}\!\sum_{{k}}\log\!\left(\!1-\!\left(\!\frac{2\sum_\mu\tilde{\Pi}_\mu
p_{\mu,{k}}}{M_{{k}}^2+\sum_\mu
(p_{\mu,{k}}^2+\tilde{\Pi}_\mu^2)}\!\!\right)^{\!\!\!2}\right)\!, \end{split} 
\eeq
where
we work in discretised Euclidean time with spacing $a_0$ and anti-periodic boundary conditions after $N_0$ time steps,
such that the momentum lies in the Brillouin zone $k=(k_0,\boldsymbol{k})\in{\rm
BZ}=[-\pi,\pi)^3$. Here, we have introduced the short-hand notation
$\sum_{{k}}=\sum_{{k}\in{\rm BZ}}/N_0N_1N_2$, $p_{\mu,{k}}=2\kappa_\mu\sin
k_\mu$, and $M_{{k}}=\tilde{m}+1-2\kappa_\mu\cos k_\mu+\tilde{\Sigma}_0$, with
$\kappa_\mu=1/2a_\mu(\sum_\nu a^{-1}_\nu)$, and  $r_\mu=1$. As customary in lattice field
theory, we work with dimensionless quantities~\cite{comment_dimensionless}, and we
set $\tilde{\Pi}_0=0$  above.

 We now minimise the effective potential considering a homogeneous scalar condensate
$\Sigma_0=\langle\sigma({x})\rangle$ $\forall x$,  and then obtain its value  by
evaluating self-consistently the corresponding tadpole diagram at the   minimum
of the effective potential 
\beq \label{eq:sigma}
\tilde{\Sigma}_0=\tilde{g}^2\sum_{{k}}\frac{M_{{k}}}{M_{{k}}^2+\sum_\mu(p_{\mu,{k}}+\tilde{\Pi}_\mu)^2}.
\eeq

\begin{figure}
\centering
\includegraphics[width=0.85\columnwidth]{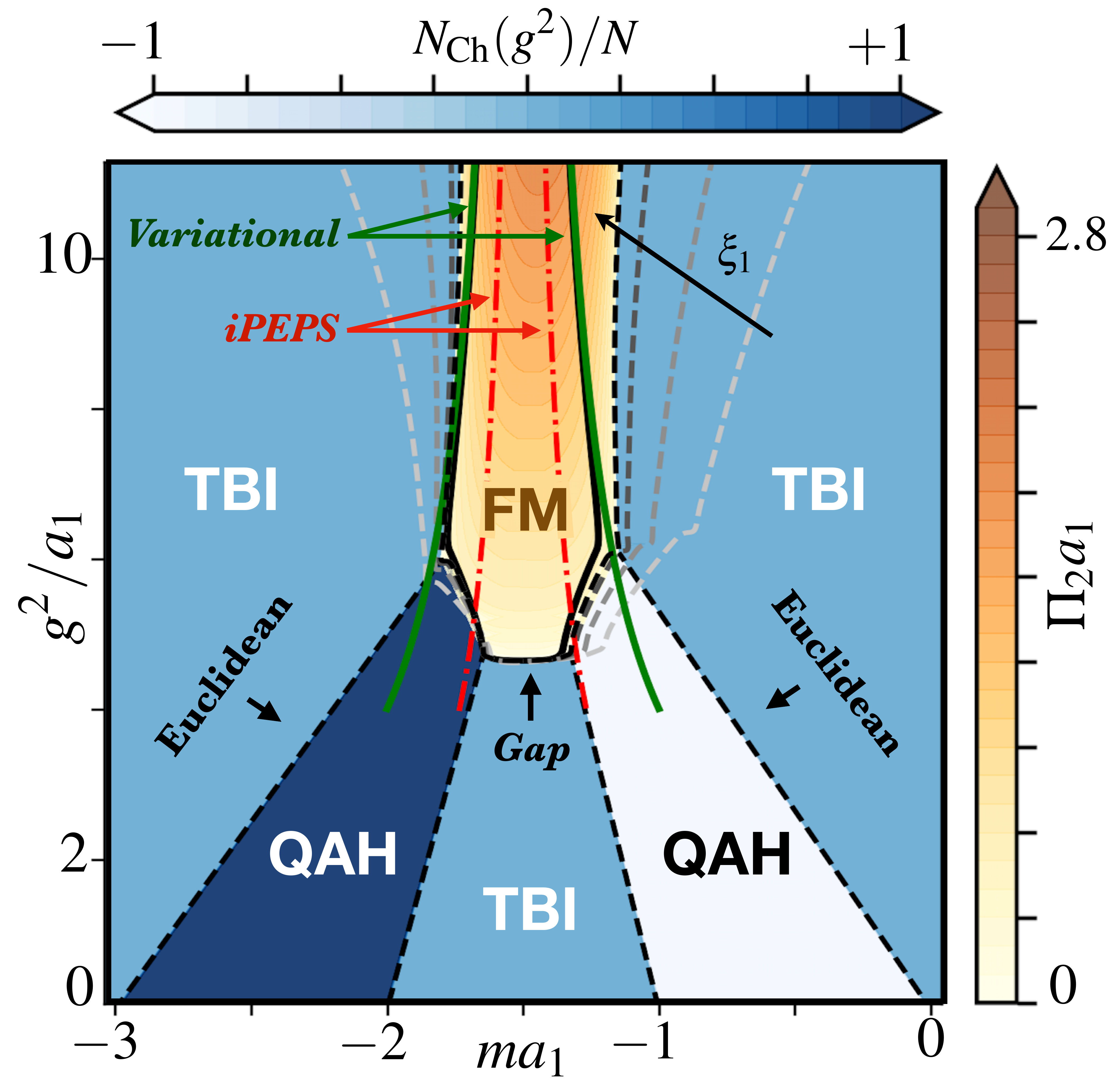}
\caption{ {\bf Phase diagram of the four-Fermi QFT:} Contour plots of the Chern number $N_{\rm Ch}$ (blue) and $\pi$ condensate (orange) for $a_2=2a_1$,  predicting a large-$N$ quantum anomalous Hall (QAH) effect, trivial band insulators (TBIs), and a ferromagnetic phase (FM). These phases are separated by dashed  lines (Euclidean) obtained from Eqs.~\eqref{eq:v_eff}-\eqref{eq:sigma} in grey scale with an increasing time-like anisotropy $\xi_1=a_1/a_0\in\{10,20,40,64\}$. The black solid line is obtained by solving self-consistent  equations (Gap)  in the time-continuum limit $a_0\to 0$, which can only delimit the area of   the FM, but give no further information about the TBI or QAH phases. The green solid lines represent the  product-state prediction for the compass model (Variational) in Eq.~\eqref{eq:compass_mf}, whereas the red dashed-dotted lines incorporate quantum correlations by the use of iPEPs. }
\label{fig_phase_diagram}
\end{figure}

		 From this pair of equations~\eqref{eq:v_eff}-\eqref{eq:sigma}, one can extract 
 all 
 effects of correlations on the large-$N$
QAH effect. First of all, in the
parity-symmetric phases where none of the $\pi$ fields condense, we find  a
non-zero scalar condensate $\Sigma_0(g^2)\neq0$ 
for any $g^2>0$~\cite{comment}. 
This condensate contributes to the static  part  of the fermionic self-energy
$\Sigma(0,\boldsymbol{k})=\Sigma_0(g^2)\gamma^0$ defined via the
inverse of the interacting Green's function  $G^{-1}(\ii k_0,\boldsymbol{k})=\ii
k_0-(M_{{k}}\gamma^0+\sum_j p_{j,{k}}\gamma^0\gamma^j)+\Sigma(\ii
k_0,\boldsymbol{k})$, which is obtained from the Fourier and Matsubara transforms of the
Euclidean two-point function $G(x-y)=\langle
\mathcal{T}\{\Psi^\dagger(x)\Psi(y)\}\rangle$. Remarkably, the static
self-energy  can be  used for a practical
calculation~\cite{PhysRevX.2.031008,PhysRevB.86.165116,Wang_2013} of
topological invariants~\cite{PhysRevLett.105.256803} beyond the non-interacting
limit~\cite{PhysRevLett.49.405}. In our case, we  obtain a closed
expression for the dependence of the Chern number on the interaction strength
\beq \label{eq:Chern_int} N_{\rm
Ch}(g^2)=\frac{N}{2}\sum_{\boldsymbol{n}_d}(-1)^{(n_{d,1}+n_{d,2})}{\rm
sign}\left(m_{\boldsymbol{n}_d}+\Sigma_0\!\!\left(g^2\right)\!\right), \eeq
which can be used to predict the existence of correlated QAH phases in which the
interacting fermions still display a quantised transverse conductivity
$\sigma_{xy}=\pm\frac{e^2}{h}N$. These  regions, depicted in Fig.~\ref{fig_phase_diagram}, 
show that the correlated Chern insulators  are very robust, surviving
for considerably large interactions with respect to the free-fermion QAH phase~\eqref{eq:Chern_free}.

		Let us now discuss the 
		$\pi$
condensation, which occurs via 
two possible channels depending on the anisotropy,
$(\Pi_1,0)$ for $a_1>a_2$, or $(0,\Pi_2)$ for $a_1<a_2$, each of which
corresponds to ferromagnetic ordering along a different axis~\cite{comment_2}. As  clarified below, 
this
ordering differs from the standard ferromagnetism found in solid-state materials
as, in particular, the low-energy excitations in this ferromagnet will be  gapped. 
The set of parameters where the $\pi$ condensates form define
critical lines that separate the  correlated QAH phase 
from a long-range-ordered ferromagnet (FM), as depicted in Fig.~\ref{fig_phase_diagram}. From a
closer analysis of the effective potential, we note that these critical lines
describe second-order quantum phase transitions which, inteerstingly, become
first-order around the central region. This  will be discussed in
detail in future work~\cite{long_version}.

		 \textit{Quantum compass model and
projected-entangled pairs.--} The above results have unveiled a rich phase
diagram with correlated Chern insulators, trivial band insulators, and 
ferromagnetic parity-broken phases, separated by critical lines related to the
specific values  of $\sigma$ and $\pi$ condensates, and controlled by strongly-coupled
QFTs. 
However, these
predictions are strictly valid in the $N\to\infty$ limit, whereas the
 cold-atom realisation of SOC~\eqref{eq:cold_atom_h}  yields $N=1$. 
  Although future experiments could validate if our large-$N$ 
predictions survive in the ultimate quantum limit of $N=1$,  exploring the  characteristic scaling of the strongly-coupled
fixed points, it would be reassuring to have a partial confirmation  with
different  methods.   
                 
		 Exploiting the connection of the four-Fermi lattice theory~\eqref{eq:H_lattice}  to the Hubbard model with SOC~\eqref{eq:cold_atom_h}, we
can derive an effective description for the strong-coupling limit $\tilde{g}^2\gg1$
using the concept of  magnetic exchange
interactions~\cite{PhysRev.79.350,ANDERSON196399}. This limit is governed by
second-order processes, by which the  fermions tunnel back and forth  forming virtual
 double occupancies. In contrast to the standard Hubbard model, where
this leads to  Heisenberg interactions~\cite{Heisenberg1928}, we  obtain 
\beq 
\label{eq:compass}
H_{\rm
eff}=\sum_{\boldsymbol{x}\in\Lambda_{\rm
s}}\left(J_x\tau^x_{\boldsymbol{x}}\tau^x_{\boldsymbol{x}+a_2\textbf{e}_2}+J_y\tau^y_{\boldsymbol{x}\phantom{+a_1\textbf{e}_1}\hspace{-10ex}}\tau^y_{\boldsymbol{x}+a_1\textbf{e}_1}-h\tau^z_{\boldsymbol{x}}\right).
\eeq 
Here, the  bi-linears
$\{\tau^\alpha_{\boldsymbol{x}}={a_1a_2\Psi}^\dagger\!(\boldsymbol{x})\sigma^\alpha\Psi(\boldsymbol{x})\}_{\alpha=x,y,z}$
yield the spin-1/2  operators  $\boldsymbol{S}_{\boldsymbol{x}}=\half\boldsymbol{\tau}_{\boldsymbol{x}}$ of the SOC model, and 
\beq
\label{eq:exchange_couplings}
J_x=-\frac{a_1}{2a_2g^2}=-\frac{2\tilde{t}_2^2}{U_{\uparrow\downarrow}},  \hspace{2ex}J_y=-\frac{a_2}{2a_1g^2}=-\frac{2\tilde{t}_1^2}{U_{\uparrow\downarrow}},
\eeq
are the exchange couplings, while $h=(m+a_1^{-1}+a_2^{-2})$ is a transverse  field. This  spin
model belongs to the family of {\it quantum compass models}~\cite{RevModPhys.87.1},
which have a characteristic   directionality of the spin-spin interactions that
is responsible for the appearance of  topologically-ordered phases of matter
with anyonic excitations in the honeycomb lattice~\cite{KITAEV20062}. For our
rectangular lattice, the anisotropic compass model has been thoroughly studied
for a vanishing transverse field $h=0$~\cite{RevModPhys.87.1}. 
In this case, there are 
 gauge-like symmetries~\cite{PhysRevB.71.024505} that
enforce a 2-fold degeneracy of the eigenstates, and can be exploited to encode
a logical qubit for fault-tolerant quantum
computing~\cite{PhysRevA.73.012340,PhysRevX.9.021041}. As one tunes the exchange
couplings across the symmetric point $J_x=J_y$, a first-order  phase transition
between two gapped ferromagnetic orders has been
detected~\cite{PhysRevB.72.024448,PhysRevB.75.144401,PhysRevLett.102.077203},
i.e. $\langle\tau^x_{\boldsymbol{x}}\rangle\neq 0$ when $J_x>J_y$, and $\langle\tau^y_{\boldsymbol{x}}\rangle\neq 0$ when $J_x<J_y$.
. 
                 
	     In contrast to the zero-field case, to the best of our knowledge,
the transverse-field quantum compass model remains largely unexplored.     Note
that the above order parameters correspond exactly to the previously introduced
$\pi$ condensates which, according to our large-$N$ results,  may  also appear for $h\neq 0$.
We have performed a variational study of the model using two different
{\em Ans\"atze}.
On the one hand, a simple separable state where all spins point  in a given
direction along the equatorial plane allows us to predict  two types of
second-order phase transitions with the following orderings 
\beq \begin{split}
\label{eq:compass_mf}
\langle
\tau^x_{\boldsymbol{x}}\rangle=\Pi_1=\left(1-\frac{h^2}{4J_x^2}\right)^{\!\!\!1/2},\hspace{2ex}
{\rm if} \hspace{1ex} 2|J_x|\geq\left|h\right|, \hspace{1ex}   J_y>J_x,\\
\langle\tau^y_{\boldsymbol{x}}\rangle=\Pi_2=\left(1-\frac{h^2}{4J_y^2}\right)^{\!\!\!1/2},\hspace{2ex}
{\rm if} \hspace{1ex} 2|J_y|\geq\left|{h}\right|,\hspace{1ex}   J_y<J_x, \end{split}
\eeq    
The groundstate displays a $\Pi_1\propto\langle\tau^x_{\boldsymbol{x}}\rangle$ $ (\Pi_2\propto\langle\tau^y_{\boldsymbol{x}}\rangle)$ condensate for $J_x<J_y<0$ $ (J_y<J_x<0)$  which, according to Eq.~\eqref{eq:exchange_couplings}, occurs for $a_1>a_2$ $ (a_1<a_2)$ in agreement  with the large-$N$ results.    These critical lines are compared to the large-$N$ prediction in Fig.~\ref{fig_phase_diagram}, showing a remarkable agreement for sufficiently strong couplings.
Finally, we have also used an infinite projected-entangled-pairs (iPEPS)
ansatz \cite{verstraete_PEPS, PhysRevLett.101.250602,PhysRevLett.101.090603} for the groundstate of Eq.~\eqref{eq:compass}.
The iPEPS is a powerful framework for the simulation of 2D strongly-correlated models, which captures the interplay of locality and entanglement
by expressing an entangled many-body wave-function in terms of local tensors.  
We have implemented 
the 
variational  iPEPS  algorithms  described in~\cite{PhysRevB.80.094403,PhysRevB.92.035142, PhysRevB.94.035133}, the  accuracy of which 
relies on a refinement parameter 
$D$ 
related to the maximum
entanglement content of the ansatz. 
In practice, increasing 
$D$ leads to better descriptions 
of the ground state, and $D=2$ is the minimum value that captures the effect of quantum correlations. 
In Fig.~\ref{fig_phase_diagram} (red  line), we present the results
for 
 $D=2$ obtained by locating the divergence of the  magnetic susceptibility $\chi_j=\partial M_j/\partial h$. 
As shown in~\cite{PhysRevLett.102.077203} for the $h=0$ model~\eqref{eq:compass}, iPEPS  with $D = 2$  already yields better results  than those obtained
by combining fermionization with mean-field theory. We have checked that, for $h\neq 0$, iPEPS also provides significantly-lower variational energies than the ones obtained by large-$N$ or the separable-state mean-field ansatz, which typically under-estimate the effect of the transverse field. This has allowed us to draw  a more accurate prediction with a clear  displacement of the critical lines and a narrower  FM region.


	 \textit{Outlook.--} We have shown that 2D  Hubbard models with  Dirac-Wilson SOC, which can be realised with neutral atoms in Raman lattices, give access to the ultimate quantum limit of strongly-coupled QFTs  hosting correlated Chern insulators and displaying a QAH effect. The framework hereby
presented can serve as the stepping stone to address open questions, such as finite fermion densities in  
search for a cold-atom realisations of fractional QAH effects~\cite{fqah,fqah_bis} and  contemporary relativistic QFTs with fractionalisation~\cite{fqhe_relativistic}.

\textit{Acknowledgements.--}
We are very grateful to Luca Tagliacozzo for fruitful discussions.
This project has received funding from the European Union's Horizon 2020 research and innovation programme under the Marie Sk\l{}odowska-Curie grant agreement No 665884, the Spanish Ministry MINECO (National Plan 15 Grant: FISICATEAMO No. FIS2016-79508-P, SEVERO OCHOA No. SEV-2015-0522, FPI), European Social Fund, Fundació Cellex, Generalitat de Catalunya (AGAUR Grant No. 2017 SGR 1341, CERCA/Program), ERC AdG NOQIA, EU FEDER, and the National Science Centre, Poland-Symfonia Grant No. 2016/20/W/ST4/00314. The work of S.J.H. was supported by STFC grant ST/L000369/1. A.B. acknowledges support from the Ram\'on y Cajal program RYC-2016-20066,  CAM/FEDER Project S2018/TCS- 4342 (QUITEMADCM),  and PGC2018-099169-B-I00 (MCIU/AEI/FEDER, UE).
     
\bibliographystyle{apsrev4-1}


\begin{thebibliography}{89}
\bibitem{comment_dimensionless}
  In Euclidean anisotropic lattices, dimensionless fields have a symmetric rescaling    $\Psi(\boldsymbol{x})=\tilde{\Psi}(\boldsymbol{x})/\sqrt{a_0a_1+a_1a_2+a_2a_0}$. This leads to the following dimensionless bare mass $\tilde{m}=ma_1/(1+\xi_1+\xi_2)$ with $\xi_1=a_1/a_0$ and $\xi_2=a_1/a_2$, and  coupling strength $\tilde{g}^2=(g^2/a_1)\xi_1\xi_2/(1+\xi_1+\xi_2)^2$. Additionally, due to the presence of additional doublers in the time direction, there is an additive renormalisation  $\tilde{m}\to \tilde{m}-\half \tilde{g}^2(1+\xi_1+\xi_2)^2/\xi_1.$ The condensates become dimensionless through $\tilde{\Sigma}_0=\Sigma_0a_1/(1+\xi_1+\xi_2)$, and $\tilde{\boldsymbol{\Pi}}=\boldsymbol{\Pi}a_1/(1+\xi_1+\xi_2)$.
\bibitem{comment_2}
At the isotropic
point $a_1=a_2=:a$, there is an additional $O(2)$ symmetry, such that the
parity-broken phase will spontaneously select the direction of the ferromagnetic
order in the $xy$-plane.  In this limit, one can also envisage the appearance of a Chern-Simons  
term which,  by its very nature, can only influence dynamics while our
large-$N$ solution focusses on static properties. The consequences of this term will be studied in the future~\cite{long_version}.
\end{thebibliography}

%

\end{document}